\documentclass[12pt,a4paper]{article}
\usepackage{amsmath, amssymb}
\usepackage{amsfonts}
\usepackage[mathscr]{eucal}
\usepackage{ascmac}
\usepackage[dvips]{graphicx}
\usepackage{subfigure}
\usepackage[charter]{mathdesign}
\usepackage{bm}
\usepackage{cite}
\usepackage{here}
\usepackage{color}
\usepackage{comment}

\usepackage[height=22.5cm,width=16.5cm]{geometry}
\begin{document}
\renewcommand{\thefootnote}{$\clubsuit$\arabic{footnote}}
\def\a{\alpha}
\def\b{\beta}
\def\c{\varepsilon}
\def\d{\delta}
\def\e{\epsilon}
\def\f{\phi}
\def\g{\gamma}
\def\h{\theta}
\def\k{\kappa}
\def\l{\lambda}
\def\m{\mu}
\def\n{\nu}
\def\p{\psi}
\def\q{\partial}
\def\r{\rho}
\def\s{\sigma}
\def\t{\tau}
\def\u{\upsilon}
\def\v{\varphi}
\def\w{\omega}
\def\x{\xi}
\def\y{\eta}
\def\z{\zeta}
\def\D{\Delta}
\def\G{\Gamma}
\def\H{\Theta}
\def\L{\Lambda}
\def\F{\Phi}
\def\P{\Psi}
\def\S{\Sigma}

\def\der{\partial}
\def\o{\over}
\def\beq{\begin{eqnarray}}
\def\eeq{\end{eqnarray}}
\newcommand{\gsim}{ \mathop{}_{\textstyle \sim}^{\textstyle >} }
\newcommand{\lsim}{ \mathop{}_{\textstyle \sim}^{\textstyle <} }
\newcommand{\vev}[1]{ \left\langle {#1} \right\rangle }
\newcommand{\bra}[1]{ \langle {#1} | }
\newcommand{\ket}[1]{ | {#1} \rangle }
\newcommand{\EV}{ {\rm eV} }
\newcommand{\KEV}{ {\rm keV} }
\newcommand{\MEV}{ {\rm MeV} }
\newcommand{\GEV}{ {\rm GeV} }
\newcommand{\TEV}{ {\rm TeV} }
\def\diag{\mathop{\rm diag}\nolimits}
\def\Spin{\mathop{\rm Spin}}
\def\SO{\mathop{\rm SO}}
\def\O{\mathop{\rm O}}
\def\SU{\mathop{\rm SU}}
\def\U{\mathop{\rm U}}
\def\Sp{\mathop{\rm Sp}}
\def\SL{\mathop{\rm SL}}
\def\tr{\mathop{\rm tr}}
\def\Dterm{_{\theta^2\bar{\theta^2}}}
\def\Fterm{_{\theta^2}}
\def\F*term{_{\bar{\theta}^2}}
\def\mi{m_{\phi}}
\def\mpl{M_{\rm pl}}

\def\IJMP{Int.~J.~Mod.~Phys. }
\def\MPL{Mod.~Phys.~Lett. }
\def\NP{Nucl.~Phys. }
\def\PL{Phys.~Lett. }
\def\PR{Phys.~Rev. }
\def\PRL{Phys.~Rev.~Lett. }
\def\PTP{Prog.~Theor.~Phys. }
\def\ZP{Z.~Phys. }


\newcommand{\TODO}[1]{{$[[ \clubsuit\clubsuit$ \bf #1 $\clubsuit\clubsuit ]]$}}
\newcommand{\kmrem}[1]{{\color{red} \bf $[[ $ KM: #1$ ]]$}}
\newcommand{\knrem}[1]{{\color{blue} \bf $[[ $ KN: #1$ ]]$}}
\newcommand{\mtrem}[1]{{\color{green} \bf $[[ $ MT: #1$ ]]$}}

\begin{titlepage}

\vskip 3cm

\begin{flushright}
UT-14-02
\end{flushright}

\vskip 3cm

\begin{center}
{\Huge \bfseries
Curvaton Dynamics Revisited\\
}

\vskip .65in
{\large
Kyohei Mukaida$^{\spadesuit}$, Kazunori Nakayama$^{\spadesuit,\diamondsuit}$ and Masahiro Takimoto$^{\spadesuit}$
}

\vskip .4in
\begin{tabular}{ll}
$^{\spadesuit}$ & \!\! {\em Department of Physics, Faculty of Science, }\\
& {\em University of Tokyo,  Bunkyo-ku, Tokyo 133-0033, Japan}\\[.5em]
$^{\diamondsuit}$ &\!\! {\em Kavli Institute for the Physics and Mathematics of the Universe, }\\
&{\em Todai Institute for Advanced Study,}\\
&{\em University of Tokyo,  Kashiwa, Chiba 277-8583, Japan}
\end{tabular}

\vskip .65in

\begin{abstract}

We revisit the dynamics of the curvaton in detail
taking account of effects from thermal environment,
effective potential and decay/dissipation rate for general field values and couplings.
We also consider the curvature perturbation
generated through combinations of various effects:
large scale modulation of the oscillation epoch, the effective dissipation rate
and the timing at which the equation of state changes.
In particular, we find that it tends to be difficult to explain the observed curvature perturbation by the curvaton mechanism
without producing too large non-Gaussianity if the curvaton energy density is dissipated through thermal effects.
Therefore, the interaction between the curvaton and light elements 
in the thermal plasma should be suppressed
in order for the curvaton to survive the thermal dissipation.

\end{abstract}


\end{center}

\end{titlepage}

\newpage

\tableofcontents

\section{Introduction}
\label{sec:intro}

Curvaton~\cite{Enqvist:2001zp} is a scalar field which is responsible for the generation of the large scale primordial 
curvature perturbation of the universe while does not contribute to the inflation.
Recent cosmological observations show that the scalar spectral index $n_s$ is in the range 
$n_s = 0.9603\pm 0.0073$~\cite{Ade:2013zuv}, showing strong evidence for the red-tilted spectrum.
In the curvaton model, the red-tilted spectrum can be explained either by the modestly steep tachyonic potential
or by the large field inflation model with $\epsilon\sim \mathcal O(0.01)$, where $\epsilon$ is the usual slow-roll parameter~\cite{Liddle:2000cg}.
A possible smoking-gun signature of the curvaton is the local-type non-Gaussianity, which is characterized by the
so-called non-linearity parameter $f_{\rm NL}$.
Although the non-Gaussianity is not found so far and $f_{\rm NL}$ is constrained severely~\cite{Ade:2013ydc},
the curvaton scenario is still a viable possibility for generating the observed curvature perturbation.

The curvaton field $\phi$ has to couple to other fields, otherwise the energy of the curvaton cannot
convert to that of radiation. To be concrete, we consider the following Lagrangian
\beq
\mathscr{L}=\mathscr{L}_\text{kin}-\frac{1}{2}m^2_{\phi}\phi^2-\lambda^2\phi^2|\chi|^2-y\phi\bar{\psi}\psi
+\mathscr{L}_{\rm higher}+\mathscr{L}_{\rm others},
\label{L_curv}
\eeq
where $\lambda$ and $y$ are coupling constants, $\mathscr{L}_{\rm kin}$ represents
canonical kinetic terms collectively, $\mathscr{L}_{\rm higher}$ denotes higher dimensional operators that
may induce (tiny) decay rate $\Gamma_{\phi}^{\rm higher}$ for $\phi$ and
$\mathscr{L}_{\rm others}$ indicates other interactions including the other light fields in thermal bath.
$\chi$ and $\psi$ are a complex scalar field and a Dirac fermion respectively and both are assumed
to be charged under some gauge group. We take $\chi$ and $\psi$ to be sufficiently light so that their zero-temperature
mass terms can be neglected in the following discussion.
The typical coupling constant of $\chi$ and $\psi$ with other lighter fields is represented by $\alpha$, which is assumed
to be relatively large: $\lambda, y \ll \alpha$.
One of the well motivated cases in our setup is the model where $\chi$ is the standard model Higgs boson.
Such a scenario was studied in Ref.~\cite{Enqvist:2013qba,Enqvist:2013gwf}.

In usual curvaton scenarios, one often assumes that the curvaton field $\phi$ starts to oscillate with
zero-temperature mass and then decays perturbatively with a constant rate. 
However, the actual evolution of the scalar field with interactions given in (\ref{L_curv}) in thermal environment
is much more complicated.
First, thermal background and/or loop corrections modify the effective potential for $\phi$, 
which change the properties of the scalar field oscillation. 
It also drastically changes the effective decay/dissipation rate of $\phi$, depending on the curvaton mass, oscillation amplitude,
coupling constants and the background temperature.
These effects on the evolution of the scalar field were studied and summarized in Refs.~\cite{Mukaida:2012qn,Mukaida:2013xxa}.
In this paper we revisit the evolution of the scalar field in general initial values, masses, couplings and temperature
in the context of curvaton scenario.
To the best of our knowledge, no literatures have considered
the curvaton scenario including effects from its interactions with thermal bath with a satisfactory level.
Actually, those effects drastically modify the evolution of the curvaton and the resulting curvature perturbation
as we will see.

This paper is organized as follows. In Sec.~\ref{sec:form} we briefly sum up ingredients needed to
follow the evolution of the universe. The list of effective potential and dissipation rates and coupled equations
of the energy components can be found in this section. 
Sec.~\ref{sec:ECP} is devoted to the formulation to evaluate the curvature perturbation.
In the formulation, we take the effects of general forms of effective potential and effective dissipation rate 
into account.
In Sec.~\ref{sec:case}, we show numerical results for some typical cases.
We conclude in Sec.~\ref{sec:conc}.

\section{Ingredients}
\label{sec:form}

\subsection{Effective potential}
\label{sec:EP}

The effective potential of $\phi$ can be approximately written as 
\beq
V(\phi)=V_{\rm tree}+V_{\rm CW}+V_{\rm thermal},
\eeq
where $V_{\rm tree}$ is the tree-level potential:
\begin{align}
	V_\text{tree}=\frac{1}{2}m^2_{\phi}\phi^2,
\end{align}
and $V_{\rm CW}$ is the Coleman-Weinberg (CW) potential \cite{Coleman:1973jx}:
\begin{align}
	V_{\rm CW}=N_{\chi}\frac{\lambda^4\phi^4}{32\pi^2}\ln\frac{\lambda^2\phi^2}{Q^2}-
	N_{\psi}\frac{y^4\phi^4}{16\pi^2}\ln\frac{y^2\phi^2}{Q^2},
\end{align}
with $N_{\chi/\psi}$ being the number of complex component of $\chi$ and the number of Dirac fermion
of $\psi$ respectively, and $Q$ being the renormalization scale.
Roughly speaking, $V_\text{CW}$ can be regarded as the quartic potential for $\phi$.
$V_{\rm thermal}$ denotes the thermal potential induced by the background thermal plasma,
which is produced from the decay of the inflaton.
It has the following form approximately:
\begin{align}
	V_{\rm thermal} \simeq&~ \theta(T-\lambda\phi)N_{\chi}\frac{\lambda^2T^2}{12}\phi^2+
	\theta(T-y\phi)N_{\psi}\frac{y^2T^2}{12}\phi^2\\
	&+\theta(\lambda\phi-T)a_{L,\chi}{\alpha^{(g)}_{\chi}}^2T^4\ln\frac{\lambda^2\phi^2}{T^2}
	+\theta(y\phi-T)a_{L,\psi}{\alpha^{(g)}_{\psi}}^2T^4\ln\frac{y^2\phi^2}{T^2},
\end{align}
where the first line represents the thermal mass term~\cite{Dolan:1973qd} which is generated if $\chi$ or $\psi$ are in thermal bath.
Here $\theta$ is the step function which approximates the effect of Boltzmann suppression.
The second line is the so-called thermal log potential~\cite{Anisimov:2000wx},
with $\alpha^{(g)}_{\chi/\psi}$ being the fine structure constant of some gauge group 
under which $\chi/\psi$ is charged. The coefficient $a_{L,\chi/\psi}$ is a model dependent order one constant and
we take $a_{L,\chi/\psi}=1$ in the following analysis for simplicity.

As the universe evolves, the temperature of thermal bath and the amplitude of $\phi$ decrease gradually.
As a result, the dominant term in the potential also varies with time.
Hence, it is convenient to define the effective mass of $\phi$ which characterizes the motion of $\phi$:
\begin{align}
	m_{\phi,\text{eff}}^2(\tilde{\phi})\equiv \max \left[m_{\phi}^2, m_{\rm CW}^2,m_{\rm th}^2,m_ {\rm th-log}^2 \right],
\end{align}
where $\tilde{\phi}$ denotes the amplitude of oscillating $\phi$.
 $m^2_{\rm CW}$ comes from $V_{\rm CW}$:
\begin{align}
	m^2_{\rm CW}=N_{\chi}\frac{\lambda^4\tilde{\phi}^2}{16\pi^2}\ln\frac{\lambda^2{\tilde{\phi}}^2}{Q^2}
	-N_{\psi}\frac{y^4\phi^2}{8\pi^2}\ln\frac{y^2\tilde{\phi}^2}{Q^2},
\end{align}
$m^2_{\rm th}$ from thermal mass:
\begin{align}
	m^2_{\rm th}=\theta(T-\lambda\tilde{\phi})N_{\chi}\frac{\lambda^2T^2}{6}+
	\theta(T-y\tilde{\phi})N_{\chi}\frac{y^2T^2}{6},
\end{align}
and $m^2_{\rm th-log}$ from thermal log term:
\begin{align}
	m^2_{\rm th-log}=\theta(\lambda\tilde{\phi}-T){\alpha^{(g)}_{\chi}}^2\frac{2T^4}{\tilde{\phi}^2}
	\ln\frac{\lambda^2\tilde{\phi}^2}{T^2}
	+\theta(y\tilde{\phi}-T){\alpha^{(g)}_{\psi}}^2\frac{2T^4}{\tilde{\phi}^2}\ln\frac{y^2\tilde{\phi}^2}{T^2}.
\end{align}
With this definition of $m^2_{\phi,{\rm eff}}$, we can write the ``velocity'' of $\phi$ near the 
origin as\footnote{
	In the case where the scalar $\phi$ oscillates dominantly with the thermal log potential,
	the very origin may be governed by the thermal mass,
	if the coupled light particles $\chi/\psi$ are produced efficiently from the thermal bath
	when the $\phi$ passes through $|\phi| < T/(\lambda~\text{or}~y)$.
	See the discussion below Eq.~\eqref{eq:t_thermal}.
	In this case, the velocity of $\phi$ near the origin is given by 
	$\left. \dot \phi \right|_{\phi \simeq 0} \simeq T^2$.
}
\begin{align}
	\left.\dot{\phi}\right|_{\phi\simeq0}\simeq \sqrt{m^2_{\phi,{\rm eff}}\tilde{\phi}^2}. 
\end{align}
The Hubble scale when $\phi$ starts to oscillate is obtained as
\begin{align}
	9H_{\rm os}^2 =\frac{\partial V(\phi_i)}{\partial \phi_i}\frac{1}{\phi_i}\sim m_{\phi\text{,eff}}^2(\phi_i),
\end{align}
where $\phi_i$ is the initial value of $\phi$.
%
%
%
%
\subsection{Effective dissipation rate}
\label{sec:ED}

After $\phi$ starts to oscillate, $\phi$ obeys the following equation:
\begin{align}
	\ddot{\phi}(t)+3H\dot{\phi}(t)+V'(\phi)+\Gamma_{\phi}[\phi(t)]\dot{\phi}(t)=0.
\end{align}
Generally, the dissipation rate $\Gamma_\phi[\phi(t)]$ depends on the field value $\phi(t)$.
In order to follow the cosmological dynamics of oscillating scalar, it is convenient to take the oscillation time average, which is 
much shorter than the cosmic time scale.
Then, the important parameter which determines the evaporation time of scalar condensation
is the oscillation averaged dissipation rate:
\begin{align}
\label{os_av_dec}
	\Gamma_{\phi}^\text{eff} (\tilde{\phi}(t))\equiv \frac{\langle\Gamma_{\phi}[\phi(t)]{\dot\phi(t)}^2\rangle}
	{\langle{\dot{\phi}}^2(t)\rangle},
\end{align}
where $\langle \cdots\rangle $ denotes the oscillation average.
The oscillating scalar $\phi$ is expected to disappear at
$H \sim \Gamma_{\phi}^\text{eff}$.
\footnote{
	An exception is the case in which the scalar shows a fixed-point behavior. See Sec.~\ref{sec:fix}.
}

In this subsection, we estimate the effective dissipation rate $\Gamma_{\phi}^\text{eff}$.
In order to evaluate $\Gamma_{\phi}^\text{eff}$, 
one has to know
the thermal mass and width of $\chi$ and $\psi$.
To avoid possible model-dependent complications,
we parametrize the thermal mass $m_{{\rm th},\chi/\psi }$ and width $\Gamma_{{\rm th},\chi/\psi}$
of $\chi/\psi$ as follows:
\beq
m_{{\rm th},\chi/\psi }&=&g_{{\rm th},\chi/\psi }T,\\
\Gamma_{{\rm th},\chi/\psi}&=&\alpha_{{\rm th},\chi/\psi}T.
\eeq
As is mentioned,
we assume that $\chi$ and $\psi$ interact with thermal bath via relatively strong couplings 
and hence the relation $\alpha_{{\rm th},\chi/\psi}\gg \lambda,y$ holds. 
We set $\alpha_{{\rm th},\chi/\psi} \sim g_{{\rm th},\chi/\psi }^2$ for simplicity.
In the following, let us consider the case where
$\chi$ and $\psi$ have decay channels if they are heavy enough,
and the decay rate is parametrized as
\begin{align}
	\Gamma_{{\rm dec}, \chi/\psi}= h^2_{\chi/\psi} m_{\chi/\psi}
\end{align}
with $h_{\chi/\psi}^2 \lesssim \alpha_{\text{th},\chi/\psi}$.
The parameters $g_{{\rm th},\chi/\psi }$, $\alpha_{{\rm th},\chi/\psi}$ and $h_{\chi/\psi}$ strongly depend on models,
and hence we regard them as free parameters in this section.
For simplicity, we assume $h_{\chi/\psi}^2 \sim \alpha_{\text{th},\chi/\psi}$ in the following.

Before moving to evaluation of averaged dissipation rate,
there is a comment on the accuracy of our estimation.
It is not easy to derive precise dissipation rates due to thermal effects, non-perturbative effects.
Therefore, we focus on the order-of-magnitude estimations and drop off numerical factors.

\subsubsection{Dissipation by non-perturbative particle production}
\label{sec:DNP}

First, let us consider the non-perturbative particle production of $\chi$ and $\psi$.
Note that the dispersion relations of $\chi$ and
$\psi$ depend on $\phi(t)$:
\begin{align}
	\omega^2_{\chi/\psi,\bm{k}}=\bm{k}^2 +m_{{\rm th},\chi/\psi }^2+(\lambda^2\text{ or } y^2)\phi^2(t).
\end{align}
where the last term $(\lambda^2 \text{ or } y^2)$ indicates $\lambda^2$ for $\chi$ and $y^2$ for $\psi$.
As $\phi(t)$ oscillates with a large amplitude, the adiabaticity of some modes may be broken down:
$|\dot{\omega}_{\chi/\psi,\bm{k}}/\omega^2_{\chi/\psi,\bm{k}}|\gg 1$.
In such a situation, the particle of $\chi/\psi$ can be produced non-perturbatively.
For detailed study of such a process, see Ref.~\cite{Kofman:1994rk}.
 
We have to comment on the effects of the quartic self interaction of $\phi$. It
can cause the non-perturbative production of $\phi$ particles
and subsequent turbulence phenomena \cite{Micha:2004bv,Berges:2013lsa}  (and references therein).
Nevertheless, we will neglect such effects as justified in App.~\ref{app:par}.
Roughly speaking, the weakness of the quartic couplings of $\phi$ enables us to neglect the effects of non-perturbative production
of $\phi$ particles as long as the quartic interaction dominantly comes from the CW correction.
More precisely, one can show that the typical scale of non-perturbative $\phi$ particle production given by
$Q \equiv \lambda \rho_\phi^{1/4}$ is much smaller than the typical interaction rate of $\chi/\psi$ with
the thermal plasma, $Q \ll \Gamma_{\text{th},\chi/\psi}$;
and that the cascade of produced $\phi$ particles toward the ultra-violet (UV) regime is driven by the interaction with
the thermal plasma via the quartic interaction $\lambda^2 \phi^2 |\chi|^2$ not by the four-point self interaction.
Therefore, the background thermal plasma see the $\phi$ field essentially as the slowly varying homogeneous field.
See App.~\ref{app:par} for details.

 The particle production of $\chi/\psi$ is characterized by the following parameter:
\begin{align}
k_{\ast, \chi/\psi}^2\equiv (\lambda \text{ or } y) \dot{\phi}|_{\phi=0}\simeq	 (\lambda \text{ or } y) m_{\phi,{\rm eff}}\tilde{\phi},
\end{align}
where 
$(\lambda^2 \text{ or } y^2)$ indicates $\lambda^2$ for $\chi$ and $y^2$ for $\psi^2$.
If the condition
\begin{align}
\label{eq:kast}
	k_{\ast, \chi/\psi}^2\gg {\rm max}\left[ m_{\phi,{\rm eff}}^2,m^2_{{\rm th},\chi/\psi }\right]
\end{align}
is met, then the following number density of $\chi/\psi$ particles are produced spontaneously
at the first passage of $\phi \sim 0$:
\begin{align}
	n_{\chi/\psi}\simeq N_{\chi/\psi}\frac{k_{\ast, \chi/\psi}^3}{4\pi^3}.
\end{align}
The first condition $k_{\ast, \chi/\psi}^2\gg m_{\phi,{\rm eff}}^2$ implies that
if the amplitude of $\phi$ is small compared with the effective mass of $\phi$, the non-perturbative particle production does not
occur. The second condition $k_{\ast, \chi/\psi}^2\gg  m_{{\rm th},\chi/\psi}^2$ implies that
if thermal mass of $\chi/\psi$ is large enough to maintain their adiabaticity, the non-perturbative particle production does not occur.

After $\chi/\psi$ is produced at the origin of $\phi$, the condensation $\phi$ increases its field value, which 
raises the effective mass of $\chi/\psi$ correspondingly.
Then, if the condition $h^2_{\chi/\psi} (\lambda \text{ or } y)\tilde{\phi}\gg m_{\phi,{\rm eff}}$ is met,
the produced $\chi/\psi$ decays well before $\phi$ moves back to the origin again.\footnote{
	Note that the parametric resonance is suppressed in this case because there are no previously produced particles
	which trigger the induced emissions.
}
We concentrate on this case in the following.
Through the decay of $\chi/\psi$, the condensation $\phi$ loses its energy~\cite{Felder:1998vq}. 
The effective dissipation rate of this process can be estimated as~\cite{Mukaida:2012qn}
\begin{align}
	\Gamma^{\rm NP}_{\phi}\simeq N_{\chi/\psi}\frac{(\lambda^2 \text{ or } y^2)m_{\phi,{\rm eff}}}
	{2\pi^4|h_{\chi/\psi} |}.
\end{align}

Typically, the non-perturbative particle production stops when $k_{\ast,\chi/\psi}$ drops down to
satisfy the condition $k_{\ast,\chi/\psi}\simeq m_{{\rm th},\chi/\psi}$. After that the dissipation of $\phi$
will be caused by thermally produced $\chi/\psi$ particles which we will discuss below.
%
%
%
\subsubsection{Dissipation by thermally produced $\chi/\psi$}
\label{sec:DTH}

Then, let us study the dissipation of $\phi$ due to thermally populated particles.
We will show neither the detailed computations nor the complete list of dissipation rate 
in the following, rather summarize the results relevant to our following discussion.
For detailed calculations and discussion, see Refs.~\cite{Mukaida:2012qn,Mukaida:2013xxa}.
See also Refs.~\cite{Yokoyama:2004pf,BasteroGil:2010pb,Drewes:2010pf}.
\\  \\
\noindent
$\bullet$ {\bf with $k_{\ast,\chi/\psi}\ll m_{{\rm th},\chi,\psi}$ and $m_{\phi,{\rm eff}}\ll\alpha_{{\rm th},\chi/\psi}T$}
 
In this region, 
the oscillating $\phi$ can be regarded as a slowly moving object in the fast interacting particles of thermal bath.
The time span $\delta_t$ in which $\chi/\psi$ can be regarded as a massless component compared to temperature
$(\lambda \text{ or } y)\phi(t)<T$
is estimated as $\delta_t\simeq T/k_{\ast,\chi/\psi}^2$.
 Within this span,
$\chi/\psi$ particles
are produced from thermal bath with a rate 
$\Gamma_{{\rm th},\chi/\psi} = h^2_{\chi/\psi}T$ at least.
Then, if the condition
\begin{align}
\label{eq:t_thermal}
	\delta_t\times \Gamma_{{\rm th},\chi/\psi}= h_{\chi/\psi}^2 T^2/k_{\ast,\chi/\psi}^2\gg 1
\end{align}
holds, one can say that $\chi/\psi$ is thermally produced with number density $n_{\chi/\psi}\sim T^3$
within the time span $\delta_t$.
Since the above inequality implies $\alpha_{\text{th},\chi/\psi} T^2 \gg k_{\ast,\chi/\psi}^2$,
the condition for the non-perturbative production given in Eq.~\eqref{eq:kast}
is violated if $m_{\phi,\text{eff}} < m_{\text{th},\chi/\psi}$.
Therefore, in this region, the $\phi$ condensation dissipates its energy dominantly 
via interactions with thermally populated $\chi/\psi$ particles.

The effective dissipation rates caused by thermally populated $\chi$ particles are the followings:
\begin{align}
\label{chi_dis}
\Gamma^{{\rm eff, slow},\chi}_{\phi}\sim N_{\chi}
\begin{cases}\cfrac{\lambda T^2}{\alpha_{{\rm th},\chi}\tilde{\phi}}\eta&\text{ for $T/\lambda\ll\tilde{\phi}$},\\[15pt]
\cfrac{\lambda^4 {\tilde{\phi}}^2}{\alpha_{{\rm th},\chi}T}&\text{ for $T<\tilde{\phi}\ll T/\lambda$},\\[15pt]
\cfrac{\lambda^4T}{\alpha_{{\rm th},\chi}}&\text{ for $\tilde{\phi}< T$},\\[15pt]
\end{cases}
\end{align}
with $\eta =  \alpha^{(g)}_{\chi/\psi}$ for thermal log,
otherwise $\eta = 1$.
And that caused by $\psi$ particles are estimated as
\begin{align}
\Gamma^{{\rm eff, slow},\psi}_{\phi}\sim N_{\psi}
\begin{cases}
\cfrac{y T^2}{\alpha_{{\rm th},\psi}\tilde{\phi}}\eta&\text{ for $T/y\ll\tilde{\phi}$},\\[15pt]
\cfrac{y^4 {\tilde{\phi}}^2}{\alpha_{{\rm th},\psi}T}&\text{ for $g_{{\rm th},\psi}T/y<\tilde{\phi}\ll T/y$},\\[15pt]
{ y^2 \alpha_{{\rm th},\chi}T}&\text{ for $\tilde{\phi}< g_{{\rm th},\psi}T/y$},\\[15pt]
\end{cases}
\end{align}
with the same definition of $\eta$.
\\ \\
$\bullet$ {\bf with $(\lambda \text{ or } y)\tilde{\phi}\ll m_{{\rm th},\chi/\psi}$}

In this region, one can safely assume that the $\chi/\psi$ particles are in the thermal bath
since the amplitude of oscillating scalar $\tilde\phi$ can be neglected.
Also, the non-perturbative production of $\chi/\psi$ does not occur as
can be seen from Eq.~\eqref{eq:kast}.
The effective dissipation rates caused by thermally populated $\chi$ particles are estimated as
		\begin{align}
			\Gamma_\phi^{\text{eff,small},\chi}& \sim N_{{\chi}}
			\begin{cases} 
				\cfrac{\lambda^4 \tilde \phi^2}{\alpha_{{\rm th},\chi} T} &\text{for}~~
				\sqrt{ \frac{\alpha_{{\rm th},\chi} T}{m_{\phi,\text{eff}}} }\, 
				T < \tilde \phi <  \frac{g_{{\rm th},\chi}T}{\lambda}\\[15pt]
				\cfrac{\lambda^4 T^2}{m_{\phi,\text{eff}}} &\text{for}~~ \tilde \phi <
				 \sqrt{ \frac{\alpha_{{\rm th},\chi} T}{m_{\phi,\text{eff}}} }\, T
			\end{cases}&\text{with}~~\alpha_{{\rm th},\chi} T \ll 
			m_{\phi,\text{eff}} \lesssim g_{{\rm th},\chi}T; \label{eq:av_diss_mid_m}
		\end{align}
		\begin{align}
			\Gamma_\phi^{\text{eff,small},\chi}& \sim N_\chi
			\begin{cases} 
				\cfrac{\lambda^4 \tilde \phi^2}{m_{\phi,\text{eff}}} &\text{for}~~T < \tilde \phi 
				<  \frac{g_{{\rm th},\chi}T}{\lambda}\\[15pt]
				\cfrac{\lambda^4 T^2}{m_{\phi,\text{eff}}} &\text{for}~~ \tilde \phi < T
			\end{cases}&\text{with}~~g_{{\rm th},\chi}T \lesssim m_{\phi,\text{eff}} < T; \label{eq:av_diss_large_m}
		\end{align}
\begin{align}
	\Gamma_\phi^{\text{eff,small},\chi}& \sim N_\chi
	\frac{\lambda^4{\tilde{\phi}}^2}{m_{\phi,\text{eff}}}~~\text{with}~~T< m_{\phi,\text{eff}}.
\end{align}
And that caused by thermally populated $\psi$ particles is the following
\begin{align}
	\Gamma_\phi^{\text{eff,small},\psi}\sim N_{\psi}
	\begin{cases}
	{y^2\alpha_{\text{th},\psi}T}& \text{for}~~m_{\phi,\text{eff}} < 2m_{\text{th},\psi},\\[15pt]
	y^2m_\psi&\text{for}~~2m_{{\rm th},\psi}<m_{\phi ,\text{eff}}.
	\end{cases}
\end{align}
Here we do not repeat the results in the case with $m_{\phi,\text{eff}} \ll \alpha_{\text{th},\chi} T$ for brevity.

The main difference between $\Gamma_\phi^{\text{eff,small},\chi}$ and $\Gamma_\phi^{\text{eff,small},\psi}$
is that the dissipation caused by $\chi$ particles becomes less and less efficient as the universe expands,
while that caused by $\psi$ particles remains since the Yukawa interaction allows the perturbative decay
of $\phi$ into $\psi$ particles.

In connection with this, in order for the $\phi$-condensation to disappear solely by the quartic interaction $\lambda^2 \phi^2 |\chi|^2$,
$H\sim\Gamma_{\phi}$ $(\tilde \phi \ll T)$ should be achieved before the temperature decreases as
$\alpha_{\text{th}, \chi} T \lesssim m_{\phi}$.
This implies that if $\lambda$ is larger than a critical value $\lambda_c$, the $\phi$-condensation
dissipates completely solely by this term. The critical value $\lambda_c$ is evaluated as~\cite{Mukaida:2013xxa}
\begin{align}
	\lambda_c\sim \left(\frac{m_{\phi}}{M_{\rm Pl}}\right)^{1/4},
\end{align}
with $M_{\rm Pl}$ being the reduced planck mass $M_{\rm Pl}\simeq 2.4\times 10^{18}~{\rm GeV}.$

Similarly, in the Yukawa case, 
the $\phi$ condensation dissipates completely by the thermal effects if the Yukawa coupling is larger than the critical value
\begin{align}
	y_c \sim \left( \frac{m_\phi}{\alpha_{\text{th},\psi} \mpl} \right)^{1/2}.
\end{align}

Note that even though the homogeneous $\phi$ condensation disappears solely by the quartic interaction $\lambda^2 \phi^2 |\chi^2|$,
the distribution of produced $\phi$ particles is still dominated by the infrared regime which is
much smaller than the temperature of thermal plasma.
Importantly, it is shown that whenever the homogeneous $\phi$ condensation can disappear completely,
the produced $\phi$ particles soon cascades toward the UV regime due to the scattering 
with the thermal plasma via the quartic interaction
and participates in the thermal plasma~\cite{Mukaida:2013xxa}.

\subsubsection{Dissipation by other effects}
\label{sec:DOE}

When $\phi$ has a large field value: $(\lambda\text{ or }y)\phi\gg T$,
a higher dimensional operator induced by integrating out $\chi$ or $\psi$
can cause the dissipation of $\phi$ condensation.
The dissipation rate is estimated as \cite{Bodeker:2006ij,Laine:2010cq,Mukaida:2012qn}
\begin{align}
	 \Gamma_{\phi}^{\text{eff,large},\chi/\psi}\simeq b\eta{\alpha^{(g)^2}_{\chi/\psi}}
	\frac{(\lambda \text{ or } y) T^2}{\tilde{\phi}}, ~~\text{for}~~ (\lambda \text{ or } y)\tilde{\phi}\gg T,
\end{align}
with $b$ being a factor to be $O(10^{-3})$.

In addition, as mentioned previously, 
we assume the higher dimensional term $\mathscr{L}_{\rm higher}$ in the Lagrangian 
which allows a perturbative decay of $\phi$ into light particles. 
The decay rate induced by this term is denoted by $\Gamma_{\phi}^{\rm higher}$.\footnote{
Since the decay products via this higher dimensional term are in the thermal bath,
	the dissipation rate can be modified in general by the thermal effects
	when the scalar field oscillates slowly compared with the typical time scale of thermal bath.
	However, in that regime, the dissipation is typically dominated by the renormalizable interaction term of $\chi/\psi$.
	The dissipation via the higher dimensional term dominates much later and hence one can neglect the thermal correction
	to this term practically.
}
%
%
%
%
%
\subsection{Evolution of the Universe}
\label{sec:EU}

In this subsection, we briefly summarize the equations which describe the evolution of the universe.
The amplitude $\tilde{\phi}(t)$ obeys the following equation of motion~\cite{Mukaida:2012qn}:
\begin{align}
	\frac{\partial \ln \tilde{\phi}(t)}{\partial t}=-\frac{1}{n_1}[n_2H+\Gamma_\phi^{\rm eff}],
\end{align}
where $\Gamma_\phi^{\rm eff}$ is the oscillation averaged dissipation rate defined in Eq.~(\ref{os_av_dec}),
$n_1$ and $n_2$ are numerical factors depending on which component dominates the universe 
and on  the effective mass of $\phi$:
\begin{align}
	\left(n_1, n_2 \right) =
	\begin{cases}
		\left( \cfrac{n+2}{2}, 3 \right) &\text{for vacuum potential with }V \propto |\phi|^n, \\[15pt]
		\left( 2, \cfrac{21}{8}~[\text{or }2] \right)
		&\text{for thermal mass in ID [or otherwise]}, \\[15pt]
		\left( 1, 9/4~[\text{or }1]  \right)
		&\text{for thermal log in ID [or otherwise]}
	\end{cases}
\end{align}
where ID stands for the inflaton-dominated era.

Until the dissipation of $\phi$ becomes comparable to the Hubble parameter,
the equations for the energy density of radiation component and that of the inflaton are given by
\begin{align}
	\frac{ \partial\rho_{\rm rad}}{\partial t}&=-4H\rho_{\rm rad} + \Gamma_{\rm I} \rho_{\rm I} + \tilde \Gamma_\phi \rho_\phi,\\
	\frac{\partial\rho_{\rm I}}{\partial t}&=-[3H+\Gamma_{\rm I}]\rho_{\rm I},
\end{align}
where $\rho_{\rm rad}$ is the energy density of radiation component, $\rho_{\rm I}$ and $\Gamma_{\rm I}$
denote energy density and decay rate of the inflaton, and
 $\rho_\phi$ represents the energy density of the condensation of $\phi$ defined as
 $\rho_\phi\equiv m_{\phi,{\rm eff}}^2(\tilde{\phi}){\tilde{\phi}}^2/2$,
 and the last term $\tilde \Gamma_\phi$ denotes the energy transportation from the curvaton to radiation.
 Practically, the last term becomes important when $\phi$ dominates the universe,
 and hence it can be expressed as $\tilde \Gamma_\phi = \Gamma_\phi^\text{eff}$.\footnote{
 	Strictly speaking, there is subtlety on the definition of energy density of $\phi$ and
	energy transportation from $\phi$ to radiation
	in the case of oscillation with thermal potential.
	However, in this case, the energy density of $\phi$ is at most that of one degree of freedom in
	thermal bath $\sim T^4$.
	Hence, it is merely a small change of $g_\ast$ and can be neglected practically within an accuracy of
	our estimation.
	In addition, in order for the curvaton $\phi$ not to produce too much non-Gaussianity,
	the energy density of $\phi$ when it disappears should nearly dominate the universe,
	that is, the curvaton has to oscillate with the vacuum potential at its decay.
	In this case, the energy density of $\phi$ and the energy transportation are nothing but
	$\rho_\phi = m_\phi^2 \tilde \phi^2/2$ and $\tilde \Gamma_\phi = \Gamma_\phi^\text{eff}$.
	Thus, we can neglect this ambiguity practically to estimate the curvature perturbation.
 }(See also Ref.~\cite{Mukaida:2012qn}.)
The energy density of the radiation $\rho_{\rm rad}$ is related to the temperature $T$ as
\begin{align}
	\rho_{\rm rad}=\frac{\pi^2}{30}g_{\ast}T^4,
\end{align}
with $g_{\ast}$ being the effective number of relativistic degrees of freedom.
We use that of the high temperature limit of the standard model: $g_{\ast}=106.75$.
The decay rate of the inflation can be expressed as:
\begin{align}
	\Gamma_{\rm I}^2\equiv \frac{g_{\ast}\pi^2T_R^4}{10 M_{\rm Pl}^2},
\end{align}
where $T_R$ is the reheating temperature of the universe. The Hubble parameter $H$ is given by
\begin{align}
	H^2=\frac{ \rho_{\phi}+\rho_{\rm rad}+\rho_{\rm I}}{3M_{\rm Pl}^2}.
\end{align}
With the equations above, we can trace the evolution of the universe.

Before closing this subsection, there is one remark.
Importantly, we assume that light particles thermalize instantaneously soon after they are
produced from the decay of the inflaton.
Otherwise the finite density correction to the dynamics of $\phi$ strongly depends on models
of reheating.
See Ref.~\cite{Harigaya:2013vwa} for the condition of instantaneous thermalization
during/after reheating via a relatively small rate of
perturbative decay.

%
%
%
%
%
%
%

\section{Curvature Perturbation}
\label{sec:ECP}

We have seen in the previous section how the scalar field $\phi$ evolves in the early universe.
Now we are in a position to
estimate cosmological parameters.
In order to evaluate cosmological parameters such as power spectrum 
$\mathscr{P}_{\zeta}$ and non-gausianity $f_{\rm NL}$ of local one,
it is convenient to use the $\delta N$ formalism
 \cite{Starobinsky:1986fxa,Sasaki:1995aw, Sasaki:1998ug,Lyth:2004gb, Lyth:2005fi}
which yields
\begin{align}
\label{pow}
	\mathscr{P}_{\zeta}&=\left( \frac{H_{\ast}}{2\pi}N_{,\phi_i}\right)^2,\\
	\frac{3}{5}f_{\rm NL}&=\frac{1}{2}\frac{N_{,\phi_i \phi_i}}{N_{,\phi_i}^2},
\end{align}
where $\zeta$ is the curvature perturbation and $\phi_i$ is the initial value of $\phi$ and we have expanded $\zeta$ 
in terms of the fluctuation of $\phi_i$ as
\begin{equation}
	\zeta = N_{,\phi_i} \delta \phi_i + \frac{1}{2}N_{,\phi_i\phi_i} \delta \phi_i\delta \phi_i + \dots.
\end{equation}
The observations show $\mathscr{P}_{\zeta}\simeq (5\times 10^{-5})^2$ and
$f_{\rm NL}=2.7\pm5.8~~ (68\%~ \text{C.L.})$~\cite{Ade:2013ydc,Ade:2013zuv}.

In our set up, the dependence of the initial value $\phi_i$ on the e-folding number $N$ from the spatially 
flat surface to the uniform density surface is not trivial. This is because the evolution and dissipation
of $\phi$ condensation are complicated compared to ordinary simple cases
where the curvaton field oscillates with quadratic potential and decays perturbatively,
since we consider the case where the curvaton directly interacts and dissipates its energy into thermal bath 
via renormalizable interactions.
	As we will see, the large scale curvature perturbation is generated through many steps.
	(1) Fluctuations of the initial field value $\phi_i$ yield fluctuations of the energy density of $\phi$ as in the ordinary curvaton model,
	(2) The oscillation epoch of $\phi$ may depend on $\phi_i$~\cite{Enqvist:2005pg,Kawasaki:2008mc,Kawasaki:2011pd},
	(3) The epoch at which equation of state of $\phi$ changes may also depend on the amplitude of $\phi$,
	(4) The effective decay/dissipation rate of $\phi$ may also depend on the amplitude of $\phi$. 
	We call this ``self-modulated reheating'', a variant type of the modulated reheating mechanism~\cite{Kofman:2003nx}.
	In general cases we have encountered in the previous section, the final curvature perturbation
	will be determined by the combination of all these effects.
In this section, we estimate $\zeta$ in general set up.

First, we assume that an entropy injection from $\phi$ condensation to the radiation occurs only once. More general cases with multi-time entropy injections will be discussed later. 
%
%
%
\subsection{Case without fixed point}

We consider the three stages of the universe separately:  the start of the oscillation of $\phi$,
just before the decay, and after the decay.
We focus on the time slicing with the constant $\rho_\phi$ surface.
We set $N_\phi$ as the e-folding number from the spatially flat surface to the constant $\rho_\phi$
surface. Then, we define $\zeta_\phi$ as
\begin{align}
	\zeta_\phi\equiv N_\phi-\bar{N}_{\phi},
\end{align}
with bar indicates a spatial average. It is conserved once the equation of state of $\phi$ is fixed.
We also consider the time slicing with the constant $\rho_{\rm oth}$ which denotes the total energy density
except $\phi$. We will refer to this surface as the uniform density slice of others (uds-o).
We define $\zeta_{\rm int}$ for this time slicing with the same way to that of $\rho_{\phi}$.
We neglect $\zeta_{\rm int}$ by assuming that the inflaton obtains negligible curvature
perturbation.
%
%
%
%
\subsubsection{After $\phi$-oscillation}

We assume $\rho_\phi\propto a^{-3(1+w_\phi^{\rm (dec)})}$ with $a$ being the scale factor of the universe.
We can express $\rho_\phi$ on uds-o in terms of constant $\bar{\rho}_\phi$ and $\zeta_\phi$: 
\begin{equation}
	\rho_\phi^{\rm (uds-o)}(\vec x) = \bar\rho_\phi e^{3(1+w_\phi^{\rm (dec)})\zeta_\phi }.
\end{equation}
Then,
\begin{equation}
	\zeta_\phi = \frac{1}{3(1+w_\phi^{\rm (dec)})}\ln\left( 1+ \frac{\delta\rho_\phi^{\rm (uds-o)}(\vec x)}{\bar\rho_\phi} \right).
	\label{zeta_phi}
\end{equation}
It is expanded as\footnote{Strictly speaking, spacial average of quadratic fluctuations such as $\delta^2_\phi$
do not vanish. However, the contributions of such non vanishing properties are negligible and we do not care 
about it.
}
\begin{equation}
	\zeta_\phi \simeq \frac{1}{3(1+w_\phi^{\rm (dec)})}\left( \delta_\phi -\frac{1}{2}\delta_\phi^2 \right),
\end{equation}
where
\begin{equation}
	\delta_\phi \equiv \frac{\delta\rho_\phi^{\rm (uds-o)}(\vec x)}{\bar\rho_\phi}.
\end{equation}

Now let us express or $\zeta_\phi$ (or $\delta_\phi$) in terms of the primordial fluctuation of $\phi$.
First note that if the Hubble parameter at the beginning of oscillation depends on $\phi(\vec x)$ itself, we have
\begin{equation}
	\rho_\phi^{\rm (uds-o)}(\vec x) = \rho_\phi^{\rm (ini)}(\vec x) \left( \frac{H_{\rm os}(\vec x)}{H^{\rm (uds-o)}} \right)
	^{-\frac{2(1+w_\phi^{\rm (osc)})}{1+w_{\rm tot}}},
\end{equation}
where $w_\phi^{\rm (osc)}$ and $w_{\rm tot}$ are the equation of state of $\phi$ just after the oscillation and 
the inflaton before the reheating ($w_{\rm tot} = 0$ if the inflaton oscillates around the quadratic potential), respectively, 
and $H_{\rm os}(\vec x)$ is the Hubble parameter at the beginning of oscillation.
If the potential of $\phi$ deviates from the quadratic one, $H_{\rm os}$ can depend on the initial field value of $\phi$.
Therefore, we obtain
\begin{equation}
	\delta_\phi = \frac{\delta\rho_\phi^{\rm (ini)}(\vec x)}{\bar\rho_\phi^{\rm (ini)}}
	- \frac{2(1+w_\phi^{\rm (osc)})}{1+w_{\rm tot}}\frac{\delta H_{\rm os}(\vec x)}{\bar H_{\rm os}},
\end{equation}
at the leading order in $\delta\rho_\phi^{\rm (ini)}$ and $\delta H_{\rm os}$.

In more general situation, the equation of state of $\phi$ may change at some epoch.
For example, it may be the case that the quartic potential dominates at first and then the quadratic term becomes dominant.
Let us suppose that $w_\phi$ changes from $w_\phi^{\rm (osc)}$ to $w_\phi^{\rm (dec)}$
at the slice $H=H_w(\vec x)$ at which $\phi (\vec x)={\rm const}$.
Then we have
\begin{equation}
	\rho_\phi^{(H=H_w)}(\vec x) = \bar \rho_\phi^{(H=H_w)} = 
	\rho_\phi^{\rm (ini)}(\vec x) \left( \frac{H_{\rm os}(\vec x)}{H_w(\vec x)} \right)
	^{-\frac{2(1+w_\phi^{\rm (osc)})}{1+w_{\rm tot}}}.
\end{equation}
By using this, we can express $\delta H_w(\vec x)$ in terms of $\delta \rho_\phi^{\rm (ini)} (\vec x)$ and $H_{\rm os}(\vec x)$ as
\begin{equation}
	 \delta_w =\delta_{H_{\rm os}}
	 - \frac{1}{a_w} \delta_{\phi_i} + \frac{1+a_w}{2a_w^2}\delta_{\phi_i}^2 - \frac{1}{a_w}\delta_{\phi_i}\delta_{H_{\rm os}},
\end{equation}
where
\begin{equation}
	\delta_w \equiv  \frac{\delta H_{w}(\vec x)}{\bar H_{w}} ,
	~~~\delta_{H_{\rm os}} \equiv  \frac{\delta H_{\rm os}(\vec x)}{\bar H_{\rm os}},
	~~~\delta_{\phi_i} \equiv \frac{\delta \rho_\phi^{\rm (ini)}(\vec x)}{\bar\rho_\phi^{\rm (ini)}},
	~~~a_w \equiv \frac{2(1+w_\phi^{\rm (osc)})}{1+w_{\rm tot}},
\end{equation}
up to the second order in these quantities.
Then, from the following equation,\footnote{
	We assume that the inflaton decays into radiation after $H=H_w$ but before $\phi$ decays.
}
\begin{equation}
	\rho_\phi^{\rm (uds-o)}(\vec x) = \rho_\phi^{\rm (ini)}(\vec x) 
	\left( \frac{H_{\rm os}(\vec x)}{H_{w}(\vec x)} \right)^{-\frac{2(1+w_\phi^{\rm (osc)})}{1+w_{\rm tot}}}
	 \left( \frac{H_{w}(\vec x)}{H^{\rm (uds)}} \right)^{-\frac{2(1+w_\phi^{\rm (dec)})}{1+w_{\rm tot}}} .
\end{equation}
we obtain $\delta \rho_\phi^{\rm (uds)}(\vec x)$ in terms of $\delta \rho_\phi^{\rm (ini)} (\vec x)$ and $\delta H_{\rm os}(\vec x)$ as
\begin{equation}
	\delta_\phi= \frac{b_w}{a_w} \delta_{\phi_i} - b_w \delta_{H_{\rm os}} 
	+ \frac{b_w(b_w-a_w)}{2a_w^2} \delta_{\phi_i} ^2 - \frac{b_w^2}{a_w} \delta_{\phi_i}\delta_{H_{\rm os}} 
	+ \frac{b_w(b_w+1)}{2}\delta_{H_{\rm os}}^2 ,
\end{equation}
where
\begin{equation}
	b_w \equiv \frac{2(1+w_\phi^{\rm (dec)})}{1+w_{\rm tot}}.
\end{equation}
More conveniently, from (\ref{zeta_phi}), $\zeta_\phi$ is expressed in non-linear form as
\begin{align}
	\zeta_\phi=\frac{b_w}{3(1+w_\phi^{\rm (dec)})} 
	\left[ \frac{1}{a_w}\ln\left(1+\delta_{\phi_i} \right) - \ln \left(1+\delta_{H_{\rm os}}\right) \right].
\end{align}
Furthermore, since $\rho_{\phi}^{\rm (ini)}$ and $H_{\rm os}$ generally scale as some powers of $\phi_i$,
we can also express $\zeta_\phi$ as
\begin{align}
\label{alwc}
	\zeta_\phi=\frac{k}{3(1+w_\phi^{\rm (dec)})}\ln 
	\left(1+\frac{\delta \phi_i}{\bar{\phi}_i} \right),
\end{align}
with $k$ being a scenario dependent constant of order unity in general.
For the quadratic potential, $k=2$. For more general cases, we summarize the calculation methods
in App. \ref{app:ESTK}.

\subsubsection{Before $\phi$-decay}

Let us take the $\phi$-decay surface, $H(\vec x) = \Gamma_\phi(\vec x)$:
\begin{equation}
	\rho_r^{\rm (dec)}(\vec x) + \rho_\phi^{\rm (dec)}(\vec x) = \bar \rho_{\rm tot}^{\rm (dec)}
	\left(1+ \frac{\delta \Gamma_\phi^{\rm (dec)}}{\bar\Gamma_\phi^{\rm (dec)}} \right)^2.
	\label{phi_dec}
\end{equation}
We define $\delta N_1$ as the e-folding number from the constant $\rho_\phi$ surface at which $\rho_\phi(\vec x)  = \bar\rho_\phi$
to the $\phi$-decay surface:
\begin{equation}
\begin{split}
	\rho_\phi^{\rm (dec)}(\vec x)  = \bar\rho_\phi^{\rm (dec)} e^{-3(1+w_\phi^{\rm (dec)}) \delta N_1} \\
	\rho_r^{\rm (dec)}(\vec x)  = \bar\rho_r^{\rm (dec)} e^{-4(\zeta_\phi+\delta N_1)}.
\end{split}
\end{equation}
By solving this equation, we obtain
\begin{equation}
	\delta N_1 = \frac{-1}{3(1+w_\phi^{\rm (dec)}) R_\phi + 4R_r }\left( 4R_r \zeta_\phi +
	\frac{2\delta \Gamma_\phi^{\rm (dec)}}{\bar\Gamma_\phi^{\rm (dec)}}  \right),
	\label{dN1}
\end{equation}
at the leading order, where
\begin{equation}
	R_\phi \equiv \left.\frac{\bar \rho_\phi}{\bar \rho_{\rm tot}} \right |_{\phi-{\rm dec}},
	~~~\left.R_r \equiv \frac{\bar \rho_r}{\bar \rho_{\rm tot}}\right |_{\phi-{\rm dec}}.
\end{equation}
They satisfy $R_\phi + R_r = 1$.

Here and hereafter, we consider a rather general form of the dissipation rate,
that is, $\Gamma_\phi$ depends on $\phi(\vec x)$ itself and the temperature $T(\vec x)$ as
\begin{equation}
	\Gamma_\phi \propto T^m \phi^n \propto a(t)^{-p}.
\end{equation}
where $p = m+qn$ with $\phi \propto a(t)^{-q}$.
$\delta\Gamma_\phi$ should be written by $\zeta_\phi$,
and hence $\delta N_1$ can be expressed solely by $\zeta_\phi$.

To do so, let us take four slices. (a) $\phi$-decay surface where $\Gamma_\phi^{(a)}(\vec x) = H^{(a)}(\vec x)$,
(b) constant $\phi$ surface where $\phi^{(b)}(\vec x) = \bar \phi$,
(c) constant $\Gamma_\phi$ surface where $\Gamma_\phi^{(c)}(\vec x) = \bar \Gamma_\phi$,
(d) uniform density surface of others where $T^{(d)}(\vec x) = \bar T$.
We define $\delta N_b (= \delta N_1)$ as the e-folding number from (b) to (a),
$\delta N_c$ as that from (c) to (b) and $\delta N_d$ as that from (d) to (c).
Note that $\delta N_c + \delta N_d = \zeta_\phi$.

We have following relations:
\begin{gather}
	\Gamma_\phi^{(a)}(\vec x)  = \Gamma_\phi^{(b)}(\vec x)e^{-p\delta N_b},   \label{a}\\
	\Gamma_\phi^{(b)}(\vec x)  = \bar\Gamma_\phi e^{-p\delta N_c},     \label{b}\\
	\bar\phi = \phi^{(c)}(\vec x) e^{-q\delta N_c},\label{c} \\
	\bar\Gamma_\phi \propto T^{(c) m}(\vec x) \phi^{(c) n}(\vec x) = {\rm const.},\label{d} \\
	\phi^{(c)}(\vec x) = \phi^{(d)}(\vec x) e^{-q\delta N_d}, \label{e} \\ 
	T^{(c)}(\vec x) = \bar T e^{-\delta N_d}. \label{f}
\end{gather}
From (\ref{a}) and (\ref{b}), we obtain
\begin{equation}
	\delta_\Gamma \equiv \frac{\delta \Gamma_\phi^{\rm (dec)}}{\bar\Gamma_\phi^{\rm (dec)}} 
	= \frac{\delta \Gamma_\phi^{(a)}}{\bar\Gamma_\phi^{(a)}} 
	= -p(\delta N_b + \delta N_c) + \frac{p^2}{2}(\delta N_b + \delta N_c) ^2.
	\label{del_Gam}
\end{equation}
From (\ref{c}) and (\ref{e}), we obtain
\begin{equation}
	\delta N_c + \delta N_d = \frac{1}{q}\ln\left( \frac{\phi^{(d)}(\vec x)}{\bar\phi} \right) = \zeta_\phi,
\end{equation}
as expected.\footnote{
	One can easily show that this $\zeta_\phi$ coincides with (\ref{zeta_phi}).
} Moreover, by using (\ref{d}) and (\ref{f}), we obtain
\begin{gather}
	\delta N_d = \frac{n}{p}\ln\left( \frac{\phi^{(d)}(\vec x)}{\bar\phi} \right) = \frac{qn}{p} \zeta_\phi ,\\
	\delta N_c = \frac{m}{qp}\ln\left( \frac{\phi^{(d)}(\vec x)}{\bar\phi} \right) = \frac{m}{p} \zeta_\phi.
\end{gather}

Substituting (\ref{del_Gam}) into (\ref{phi_dec}) and solving it self-consistently order by order, we obtain
\begin{equation}
	\delta N_1^{(1)} = \frac{4R_r - 2m}{2p - 3(1+w_\phi^{\rm (dec)})R_\phi - 4R_r }\zeta_\phi,
\end{equation}
at the leading order and 
\begin{equation}
	\delta N_1^{(2)} = \frac{-2R_rR_\phi}{\left[2p - 3(1+w_\phi^{\rm (dec)})R_\phi - 4R_r\right]^3 }
	\left[ 4(p-m) - 3(1+w_\phi^{\rm (dec)})(2-m) \right]^2
	\zeta_\phi^2,
\end{equation}
at the second order in $\zeta_\phi$.

\subsubsection{After $\phi$-decay}

Finally, let us take the uniform density surface after $\phi$-decay, where only the radiation exists.
\begin{equation}
	\rho_r^{(\rm uds)} (\vec x) = \bar\rho_r = 3\bar H^{\rm (uds) 2} M_P^2.
\end{equation}
We define $\delta N_2$ as the e-folding number from the $\phi$-decay surface to the uniform density surface:
\begin{equation}
\begin{split}
	\bar \rho_r = \rho_r^{\rm (dec)}(\vec x) e^{-4\delta N_2} = 3H^{\rm (dec)^2} M_P^2e^{-4\delta N_2}
	= \bar \rho_r \left(1+ \delta_\Gamma \right)^2e^{-4\delta N_2}.
\end{split}
\end{equation}
Hence we obtain
\begin{equation}
	\delta N_2 = \frac{1}{2}\ln \left(1+ \delta_\Gamma \right)
	= \frac{1}{2}\left(\delta_\Gamma - \frac{1}{2}\delta_\Gamma^2\right) .
\end{equation}
Then from (\ref{del_Gam}), we find
\begin{equation}
	\delta N_2 = -\frac{1}{2}(m\zeta_\phi + p\delta N_1).
\end{equation}

\subsubsection{Curvature perturbation}

The curvature perturbation $\zeta$ evaluated well after the decay of $\phi$ is equal to the e-folding number
from the spatially flat surface to the uniform density surface, according to the $\delta N$ formalism.
The resulting curvature perturbation is thus given by
\begin{equation}
	\zeta = \zeta_\phi + \delta N_1 + \delta N_2.
\end{equation}
At the leading order, it is given by
\begin{equation}
	\zeta^{(1)} = \frac{R_\phi}{3(1+w_\phi^{\rm (dec)})R_\phi + 4R_r}\left[
		3(1+w_\phi^{\rm (dec)}) \zeta_\phi + (3w_\phi^{\rm (dec)} -1) \frac{\delta\Gamma_\phi^{\rm (dec)}}{2\bar\Gamma_\phi^{\rm (dec)}}
	\right].
\end{equation}
It is seen that the curvature perturbation is proportional to $R_\phi$
and the effect of self-modulated reheating vanishes for $w_\phi^{\rm (dec)} =1/3$, as expected.

By substituting (\ref{del_Gam}), we obtain
\begin{equation}
	\zeta^{(1)} = \frac{R_\phi \left[ 4(p-m)-3(1+w_\phi^{\rm (dec)})(2-m) \right]}{2\left [2p - 3(1+w_\phi^{\rm (dec)})R_\phi - 4R_r \right]}
	 \zeta_\phi,
\end{equation}
at the leading order, and
\begin{equation}
	\zeta^{(2)} = \frac{(p-2)R_rR_\phi}{\left [2p - 3(1+w_\phi^{\rm (dec)})R_\phi - 4R_r \right]^3}
	\left[4(p-m)-3(1+w_\phi^{\rm (dec)})(2-m) \right]^2
	\zeta_\phi^2,
\end{equation}
at the second order in $\zeta_\phi$.

Now, let us evaluate power spectrum 
$\mathscr{P}_{\zeta}$ and non-gausianity $f_{\rm NL}$ in terms of $p,m,w_{\phi}^{({\rm dec})},k$ which
is defined in (\ref{alwc}) and $r$, that is defined as
\begin{align}
	r\equiv 
	\frac{R_\phi \left[ 3(1+w_\phi^{\rm (dec)})(2-m)-4(p-m) \right]}
	{ 2\left[(3(1+w_\phi^{\rm (dec)})-2p)R_\phi +(4-2p)R_r\right] }.
\end{align}
The power spectrum can be written as
\begin{align}
	\mathscr{P}_{\zeta}=\left(
	\frac{H_{\ast}}{2\pi \phi_i}\frac{kr}{3(1+w_\phi^{\rm (dec)})}
	\right)^2.
\end{align}
The non-linearity parameter $f_{\rm NL}$ has the following form
\begin{align}
	f_{\rm NL}=A^{(0)}+A^{(1)}r+A^{(-1)}\frac{1}{r},
\end{align}
with
\begin{align}
	A^{(0)}&=-\frac{5}{3}\times \frac{6(1+w_\phi^{({\rm dec})})-4-2p}{2-p},\\
	A^{(1)}&=-\frac{5}{6}\times \frac
	{4(4-3(1+w_\phi^{({\rm dec})}))(3(1+w_\phi^{({\rm dec})})-2p)}
	{(2-p)(3(1+w_\phi^{({\rm dec})})(2-m)-4(p-m))}
	,\\
	A^{(-1)}&= \frac{5}{4}\times\left[
	\frac{2(3(1+w_\phi^{({\rm dec})})(2-m)-4(p-m))}{3(2-p)}-\frac{2(1+w_\phi^{({\rm dec})})}{k}
	\right].
\end{align}
In most cases including the cases which we deal with, $A^{(-1)}$ is a factor of order unity, 
hence we have $f_{\rm NL} \sim 1/r$ for $r \ll 1$ as in the ordinary curvaton model.
However, there may exist a situation in which $A^{(-1)}=0$ keeping $\zeta \sim 5\times 10^{-5}$, that relaxes the constraint on
a curvaton scenario drastically.
%
%
%
%
%
%
\subsection{Case with fixed point}   \label{sec:fix}

In general, the dissipation rate of $\phi$ depends on $\tilde{\phi}$ and $T$.
There are some cases in which the dissipation becomes ineffective before the $\phi$ condensation disappears.
For example, the dissipation rate caused by $\chi$ particles may have the form given in (\ref{chi_dis}).
If the dissipation rate becomes comparable with $H$ during $\tilde{\phi}\gg T/\lambda$,
$\tilde\phi$ will soon decrease to $\tilde\phi\sim T/\lambda$. 
After that, the dissipation becomes ineffective if $\phi$ oscillates with the zero-temperature mass.
Eventually, $\phi$ disappears via the dissipation/decay.
For such a case, we can define $r$ for the each epoch at which $\Gamma \sim H$
and the dominant contribution to the curvature perturbation comes from the epoch of dissipation with maximal $r$.

There are some exceptions in which the dissipation fixes $\tilde{\phi}$ to a certain value depending only on $T$. 
For example, if $\phi$ oscillates with quartic potential after the dissipation caused by $\chi$ particles
makes $\tilde{\phi}\sim T/\lambda$, the $\tilde{\phi}$ will be fixed to $H\sim \Gamma$.
This is because the dissipation rate behaves $\Gamma\propto \tilde{\phi}^2T^{-1}$, which
drops down more slowly than $H$ if $\phi$ oscillates with quartic potential. 
In such a case, the curvature perturbation is determined at the time when $\phi$ is fixed.
This is true even if $\phi$ dominates the universe after the time of fixing.
We will call this point as fixed point.
As you will see in the section \ref{sec:case}, $\phi$ can be actually trapped at the fixed point
in the region of relatively high $T_R$ and $\phi_i$. 
%
%
%
%
%
%
%
%
\section{Case Study}
\label{sec:case}

In this section, we consider the viability of the curvaton scenario taking all the effects described so far into account.
In particular, we estimate $R$ defined below for some typical cases,
\begin{align}
\label{RRR}
	\tilde R\equiv \left. \frac{\rho_{\phi}}{\rho_{\rm rad}}\right|_{\phi\text{ disappears }},
\end{align}
which characterizes the properties of $\phi$ as a curvaton.
Once we assume that the curvaton is a dominant source of the observed density perturbation,
we need $\tilde R \gtrsim 0.1$ in order to avoid too large non-Gaussianity.
In addition, we will check whether $\phi$ comes to be trapped at the fixed point or not.
If $\phi$ is once trapped at the fixed point, generally $\tilde R \ll1$ holds at the time of fixing.
Therefore, such a region is not allowed.

In order for the curvaton field $\phi$ to generate the observed power spectrum $\mathscr{P}_{\zeta}$ in Eq.~(\ref{pow}),
the Hubble parameter during inflation, $H_{\ast}$, is fixed to be
\begin{equation}
\label{cnd:H1}
	H_{\ast}\sim 6\times 10^{-4} \phi_i \left(\frac{1+\tilde R}{\tilde R}\right) \lesssim 10^{14}\,{\rm GeV},
\end{equation}
where the second inequality comes from the condition that the amplitude of tensor mode should not be too large.
On the other hand, $H_{\ast}$ must be greater than $H_{\rm os}$, which is the Hubble parameter at the onset of
$\phi$ oscillation, otherwise $\phi$ starts to oscillate during or before the inflation.
Thus, we impose a following condition:
\begin{align}
\label{cnd:H2}
	H_{\rm os}<6\times 10^{-4} \phi_i \left(\frac{1+ \tilde R}{\tilde R}\right).
\end{align}
We have checked these conditions (\ref{cnd:H1},\ref{cnd:H2}) in the numerical study. 
%
%
\subsection{$y=0$ case}
\label{sec:case1}

First, we consider the situation in which there are no effects from fermion $\psi$ i.e.~$y=0$.
In such a case, the scalar condensation $\phi$ can not completely disappear
if the coupling $\lambda$ is smaller than the critical value $\lambda_c$
because $Z_2$ symmetry forbids the perturbative decay of $\phi$.
Therefore, we assume non-zero $\Gamma_\phi^{\rm higher}$ for $\phi$ to obtain small but nonzero perturbative decay rate. 
In order to see the typical situation, we assume the following form of $\Gamma_\phi^{\rm higher}$
\begin{align}
	{\Gamma^{\rm higher}_\phi}\equiv \sqrt[]{\mathstrut \frac{g_{\ast}\pi^2T_{\rm dec}^4}{10 M_{\rm Pl}^2}},
\end{align}
with a decay temperature $T_{\rm dec}\sim \mathcal O(1)\,\text{MeV}$. This ensures
that $\phi$ condensation decays before Big-Bang Nucleosynthesis (BBN).
Thus, for small $\lambda$ in which $\phi$ condensation cannot be dissipated away,
our calculation gives the upper bound to $\tilde R$. 
We take following values for other parameters:
$g_{{\rm th},\chi}=0.5,\alpha_{{\rm th},\chi}=0.05,\alpha_{\chi}^{(g)}=0.05,h_{\chi}=1,N_\chi=1$.
For the renormalization scale $Q$, we take $Q=100~{\rm GeV}$.
This set up is close to the minimal higgs curvaton model~\cite{Enqvist:2013gwf}.
The remaining parameters are the reheating temperature of the universe $T_R$, the tree-level mass
$m_{\phi}$ and the coupling constant $\lambda$.

\begin{figure}[t]
 \centering
\includegraphics[width=10cm,clip]{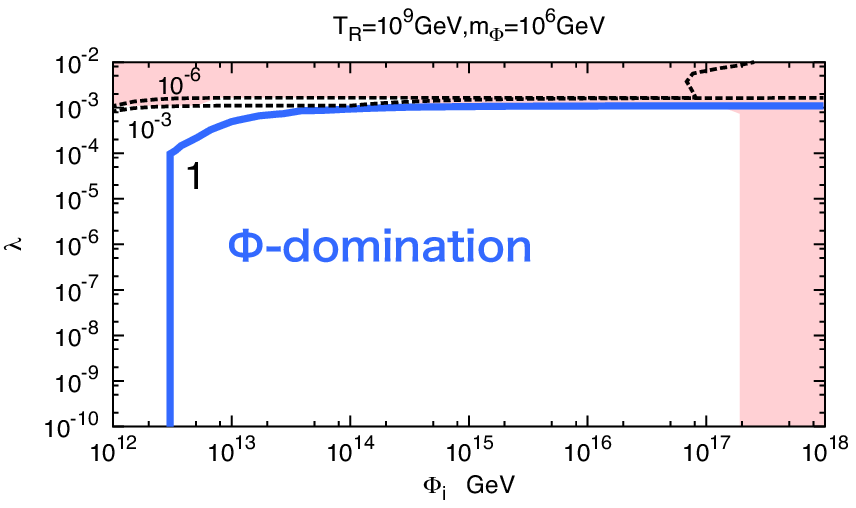}
\includegraphics[width=10cm,clip]{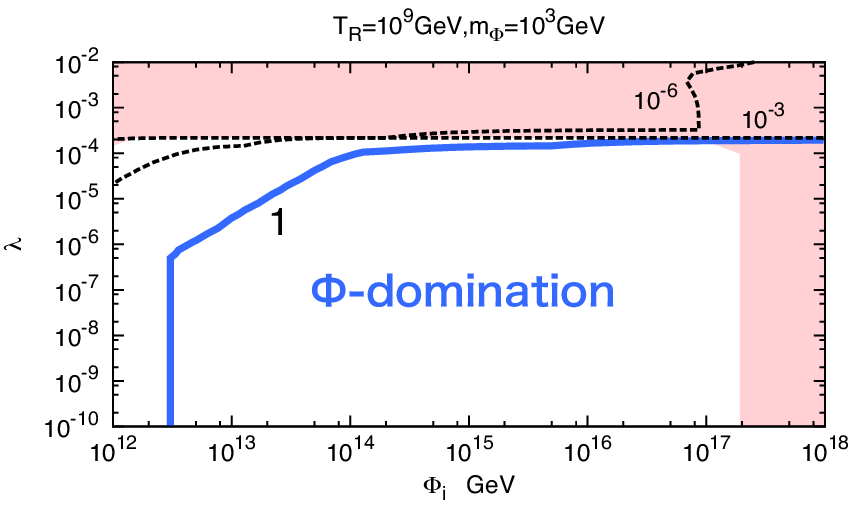}
\includegraphics[width=10cm,clip]{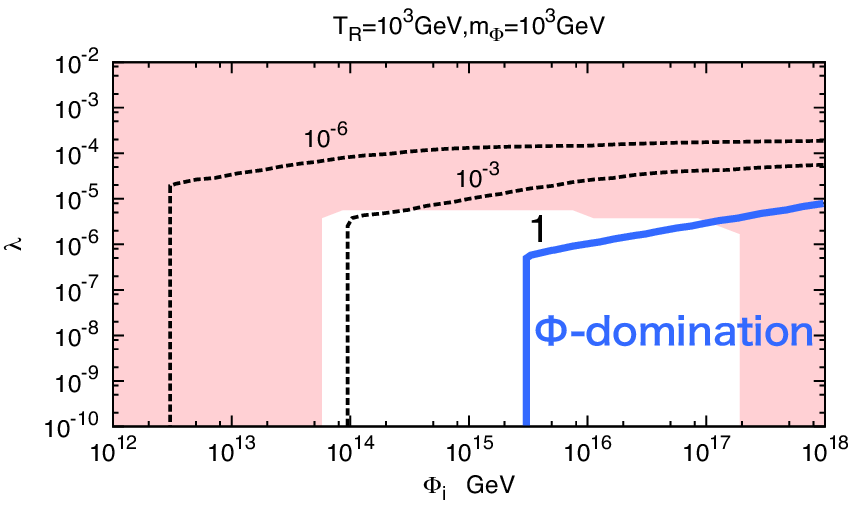}
\caption{Contours of constant $\tilde R$ on $(\phi_i,\lambda)$ plane.
  We take $(m_{\phi},T_{\rm R})=(10^3\,{\rm TeV},10^{9}\,{\rm GeV})$ (top), $(m_{\phi},T_{\rm R})=
  (1\,{\rm TeV},10^9\,{\rm GeV})$
  (middle), $(m_{\phi},T_{\rm R})=(1\,{\rm TeV},10^3\,{\rm GeV})$ (bottom). 
  In the pink shaded region, the condition (\ref{cnd:H1}) is violated.}
  \label{fig:y=0}
\end{figure}

Fig.~\ref{fig:y=0} shows contours of $\tilde R$ on $(\phi_i,\lambda)$ plane.
We take $(m_{\phi},T_{\rm R})=(10^3\,{\rm TeV},10^{9}\,{\rm GeV})$ (top), $(m_{\phi},T_{\rm R})=
  (1\,{\rm TeV},10^9\,{\rm GeV})$
  (middle), $(m_{\phi},T_{\rm R})=(1\,{\rm TeV},10^3\,{\rm GeV})$ (bottom).
  In the pink shaded region, the condition (\ref{cnd:H1}) is violated. 
  
  One can see that $\tilde R\sim1$ can be realized just below the line of $\lambda\sim\lambda_c$.
  This fact is easily understood because if $\lambda>\lambda_c$, the condensation $\phi$
  is dissipated and $\tilde R$ becomes suppressed, while if $\lambda\ll\lambda_c$, the
  condensation survives until it decays via the higher dimensional term. 
  Therefore, the line of $\tilde R\sim 1$ exists just below $\lambda=\lambda_c$.
  The difference between top and middle figures mainly comes from the position of the line $\lambda=\lambda_c$.
   For smaller $ T_{\rm R}$, $\phi$ oscillates in the inflaton dominant era for a long time and $\tilde R$ tends to be smaller. 
   The fixed point behavior is not realized in the parameter regions we considered.
%
%
%
%
%
%
%
%
%
%
%
\subsection{$y\neq 0$ case}
\label{sec:case 2}

Now we consider more general case i.e.~$y\neq 0$.
We take following values for the parameters:
$g_{{\rm th},\chi/\psi}=0.5, \alpha_{{\rm th},\chi/\psi}=0.05, \alpha_{\chi/\psi}^{(g)}=0.05, h_{\chi/\psi}=1, N_\chi=2, N_\psi=1$.
For the renormalization scale $Q$, we choose $Q=100\,{\rm GeV}$.
If we set $\lambda=y$, the quartic potential of $\phi$ coming from the CW correction vanishes 
and we call this supersymmetric (SUSY) case. 
If $y>\lambda$, the quartic potential of $\phi$ becomes unstable at the large field value and we do not consider such a case.
We take $\Gamma_\phi^{\rm higher}=0$ for simplicity, because 
the non-zero Yukawa coupling $y$ typically dominates the perturbative decay rate of $\phi$ unless $y$ is extremely small.

\begin{figure}[h]
 \centering
\includegraphics[width=8cm,clip]{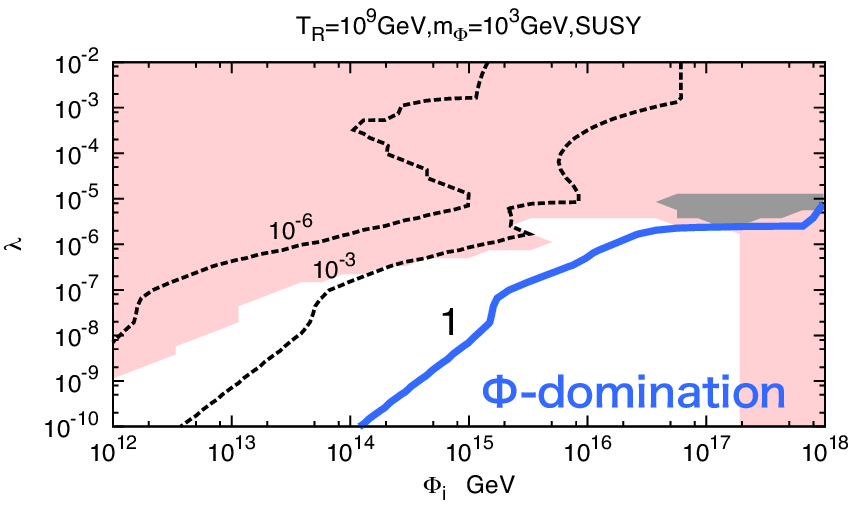}
\includegraphics[width=8cm,clip]{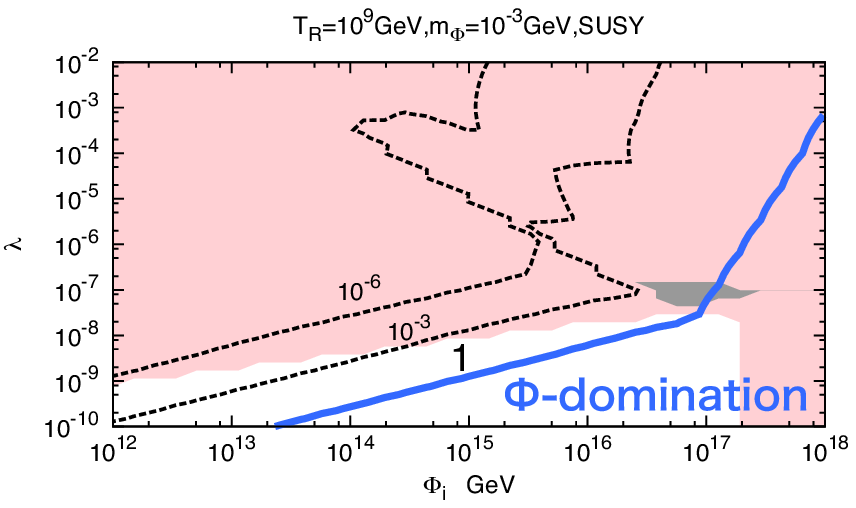}
   \caption{Contours of constant $\tilde R$ on $(\phi_i,\lambda)$ plane in SUSY case $\lambda=y$.
  We take $(m_{\phi},T_{\rm R})=(1\,{\rm TeV},10^9\,{\rm GeV})$ (left) and 
  $(10^{-6}\,{\rm TeV},10^9\,{\rm GeV})$ (right).
  In the pink shaded region, the condition (\ref{cnd:H1}) is satisfied. 
  The gray region shows where the fixed point phenomenon is realized.}
  \label{fig:SU}
  \end{figure}

Fig.~\ref{fig:SU} shows contours of $\tilde R$ in the SUSY case $\lambda=y$.
 We take $(m_{\phi},T_{\rm R})=(1\,{\rm TeV},10^9\,{\rm GeV})$ (left) and 
  $(10^{-6}\,{\rm TeV},10^9\,{\rm GeV})$ (right).
 In the pink shaded region, the condition (\ref{cnd:H1}) is violated. 
 The gray region shows where fixed point phenomenon is realized,
 although the most of these regions violate the condition (\ref{cnd:H1}).
 Since the Yukawa coupling induces the earlier dissipation/decay compared with the $y=0$ case,
 the energy fraction $\tilde R$ tends to be smaller.
 Hence the constraints become severer.
 The contours of $\tilde R$ in the figure are relatively curved in the upper side
 and tend to have large values compared with the previous case $y=0$.
 This is because thermal potential is more likely to affect
 the dynamics of the condensation of $\phi$ in the SUSY case
 and because the absence of four point  self interaction delays the beginning of oscillation,
 respectively.
 As in the previous case, above the critical coupling $y_c$,
 the curvaton dissipates its energy thermally.
 For a relatively smaller $m_{\phi}$, the fixed point phenomena tend to be realized as
 the thermal potential dominates the dynamics.
 
\begin{figure}
\centering
\includegraphics[width=10cm,clip]{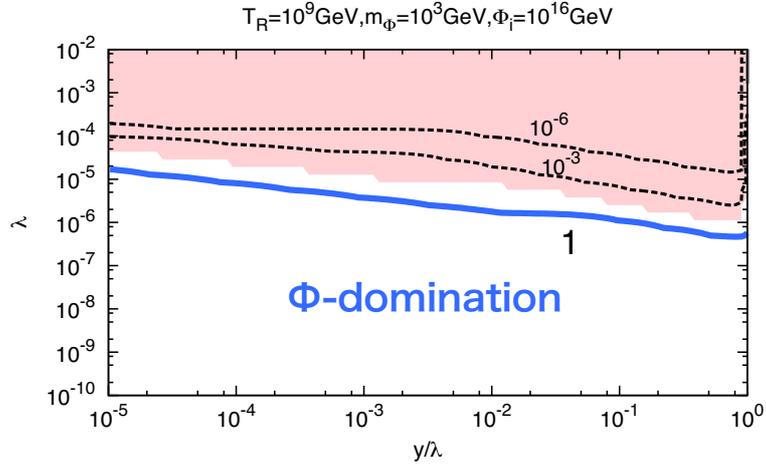}
   \caption{Contours of constant $\tilde R$ on $(\lambda,y/\lambda)$ plane with $\phi_i=10^{16}\,{\rm GeV}$.
 We have taken $(m_{\phi},T_{\rm R})=(1\,{\rm TeV},10^9\,{\rm GeV})$.}
  \label{fig:GE}
\end{figure}

 Fig.~\ref{fig:GE} indicates contour of $\tilde R$ for general set of $(\lambda,y)$ with with $\phi_i=10^{16}\,{\rm GeV}$.
 We set $(m_{\phi},T_{\rm R})=(1\,{\rm TeV},10^9\,{\rm GeV})$.
 Similar to the case of $y=0$, the line of $\tilde R=1$ lies a bit below $\lambda=\lambda_c$.
This figure indicates that the SUSY effects appear at the vicinity of $y=\lambda$. 


\clearpage

\section{Summary}
\label{sec:conc}

In this paper we have revisited the curvaton model taking account of the curvaton interactions with other light species.
The curvaton dynamics can be drastically modified compared with the ordinary scenario
in which the curvaton oscillates with quadratic potential and decays perturbatively at the late epoch.

Most importantly, the curvaton energy density at its decay/dissipation strongly depend on the curvaton initial field value, 
potential, coupling constants and background temperature.
Moreover, in general, the resulting curvature perturbation is complicated because:
\begin{itemize}
\item The oscillation epoch of $\phi$ may depend on $\phi_i$.
\item The epoch at which equation of state of $\phi$ changes may also depend on the amplitude of $\phi$.
\item The effective decay/dissipation rate of $\phi$ may also depend on the amplitude of $\phi$. 
\end{itemize}

We have considered all these effects and derived the cosmological evolution of the curvaton and its viability to explain the
observed curvature perturbation of the universe.
As is well known, in order to avoid too large non-Gaussianity, the curvaton energy fraction at its decay epoch must be close to one.
This means that the curvaton is not likely dissipated by thermal effects.
In other words, the curvaton should survive the thermal dissipation,
and hence the interaction between the curvaton and thermal plasma via the renormalizable quartic/Yukawa terms
should be suppressed.
In fact, it is shown that there is an upper bound on the renormalizable coupling, $y$ or $\lambda$,
from the constraints on the non-Gaussianity and the tensor mode.

\section*{Acknowledgment}

This work is supported by Grant-in-Aid for Scientific
research from the Ministry of Education, Science, Sports, and Culture
(MEXT), Japan, No.\ 21111006 (K.N.), and No.\ 22244030 (K.N.).
The work of K.M. and M.T. are supported in part by JSPS Research Fellowships
for Young Scientists.
The work of M.T. is supported by the Program for Leading Graduate
Schools, MEXT, Japan.


\appendix
\section{Effects of $\phi$ particles via four-point interaction}
\label{app:par}

$\phi$ particles can be generated in the early universe 
due to the non-perturbative effects caused by oscillation of $\phi$ condensation itself via the four point self interaction of $\phi$.
In this Appendix, we show such produced $\phi$ particles are harmless in our case, 
i.e.~they do not much affect the results obtained in Sec.~\ref{sec:case}. 

Concretely, we verify two things:
\begin{itemize}
\item The non-perturbative production of $\phi$ particles do not change the order-of-magnitude estimation of the amplitude of $\phi$ condensation.
\item The number density of $\phi$ particles obeys an equation similar to that of condensation.
\end{itemize}
%
%
%
%
%
\subsection{Non-perturbative production of $\phi$ particles and turbulence}

Aside from the quartic (Yukawa) interaction $\lambda^2 \phi^2 |\chi|^2$ ($y \bar \psi \psi$) in Eq.~\eqref{L_curv},
there is a possible source that drives the $\phi$ condensation towards a higher momentum,
that is, the four point self interaction of $\phi$ itself induced by the one loop effect.
In order to compare the dynamics via four point interaction with that of quartic/Yukawa one discussed in Sec.~\ref{sec:ED},
let us clarify its effect and typical time scale.
For that purpose, we consider the following potential motivated by the CW potential:
\begin{align}
	V = \lambda^4\phi^4.
\end{align}
Here we omit the loop factor for simplicity.\footnote{
 	We assume $\lambda \gg y$ to stabilize the effective potential.
}
We assume $\lambda$ is relatively small such that $\alpha\gg \lambda$ with the thermal width of
$\chi$ being $\Gamma_{\text{th},\chi}=\alpha T$ if $\chi$ participates in the thermal plasma.

In the following discussion, we assume
$\phi$ has initially large amplitude.
Such a case is well studied for example in \cite{Micha:2004bv,Berges:2013lsa}
and we follow their discussion.
According to \cite{Berges:2013lsa} , the typical scale $Q$ is written as
\begin{align}
	Q\equiv \lambda(\rho_{\phi})^{1/4}.
\end{align}
Since the Floquet index is roughly given by $\mu \sim Q$, 
the non-perturbative production of $\phi$ particles occurs during $t \lesssim Q^{-1}\log\lambda^{-4}$,
and then energy density of them becomes compatible to that of condensation at $t_\text{NP} \sim Q^{-1}\log\lambda^{-4}$.
At that time, the amplitude of $\phi$ is changed by factor not order.

After that, the distribution function obeys the self-similar behavior which is referred to as turbulent phenomena.
The distribution function of high momentum mode with $p\gg Q$ and low momentum mode with
$p\ll Q$ evolve in different ways, which is dubbed as a dual cascade~\cite{Berges:2013lsa}
and characterized by different exponents $\kappa$ of distribution function, $f(p) \propto (Q/p)^\kappa$.
\begin{itemize}
\item For low momentum modes ($p \ll Q$), 
an inverse particle cascade toward infrared takes place, which is driven by the number conserving interactions
among the soft sector ($p \ll Q$).\footnote{
Though the exact zero mode will decay by a power law with $\phi_0(t)\sim Q(Qt)^{-1/3}$~\cite{Micha:2004bv},
the particles are still condensed in low momenta regime and phenomenological consequence
is not clear. Therefore, we simply regard such a condensation below $p \ll Q$ as a zero mode effectively
in the following.
In addition,
even if the effective zero mode may decay with this power low,
the change of exponent from $\tilde \phi \propto a^{-1}$ can be neglected practically in our case
since the scalar disappears before this difference becomes significant.
}
The exponent is given by $\kappa_\text{M} = 4/3$ for $f(p)\lesssim  \mathcal{O}(1/ \lambda^4)$.
In terms of perturbative kinetic picture in $\lambda$,
the stationary particle flow driven by the four point interaction implies this exponent~\cite{Micha:2004bv}.
For ultra-soft modes $f(p) \gtrsim \mathcal{O}(1/ \lambda^4)$,
the exponent is turned out to be more stronger: $\kappa_\text{S} = 4$~\cite{Berges:2013lsa}.
This is because the perturbative expansion in $\lambda$ is broken down
and we have to consider many other processes non-perturbatively.
In the case of $N \geqslant 2$ with O($N$)-symmetric scalar field theory,
in terms of the $1/N$ expansion, it is shown that the anomalous exponent $\kappa_\text{S} = 4$ can be understood
as the consequence of momentum dependent effective coupling $\lambda_\text{eff} (p) \sim p^2$~\cite{Berges:2013lsa}.
\\
\item For high momentum modes ($p \gg Q$),
an energy cascade toward ultraviolet (UV) takes place,
which is driven by the effective three point interaction for hard modes: $\phi (\text{hard}) 
+ \phi (\text{hard}) \to \phi(\text{soft}) + \phi (\text{hard})$.
It is characterized by the Kolmogorov exponent $\kappa_\text{H} = 3/2$ and
the stationary energy flow towards UV implies this exponent~\cite{Micha:2004bv}.
The distribution function obeys the following self-similar evolution:
$f(t,p) = (Qt)^{\alpha} f_s((Qt)^\beta p)$ with $\alpha = 4\beta$ and $\beta = 1/(2n-1)$ for a $n$-point interaction.
Hence one finds $(\alpha, \beta) = (4/5,1/5)$ in this case.
The maximum momentum $p_{\rm max} \equiv Q (Qt)^\beta$
which has dominant energy density
grows higher and higher.
\end{itemize}

After the distribution function reaches $f \sim 1$,
the quantum effects become important and lead to thermal equilibrium with the Bose-Einstein distribution.
Therefore, the time scale when the turbulent phenomena stops can be estimated as
\begin{align}
	t_{\rm quant}\sim Q^{-1}\lambda^{-5}=\lambda^{-6}(\rho_\phi)^{-1/4},
\end{align}
since $f_s \sim \mathcal{O}(1/\lambda^4)$.
At that time, the maximum momentum arrives at $p_\text{max} \sim \rho^{1/4}_\phi > Q$.

Now we are in a position to discuss the effects of four point interaction on the dissipation caused by the quartic/Yukawa interaction.
To maximize the effect of four-point interaction, let us concentrate on the case with $\lambda^2 \phi_i \gg \lambda T$.

First, the non-perturbative production of $\chi$ particles may take place as discussed in Sec.~\ref{sec:DNP},
and it terminates at $k_\ast^2 \sim \lambda^3 \tilde \phi^2 \sim \alpha T^2$.
At that time,
the ratio of energy density of $\phi$ to that of background thermal plasma is given by
\begin{align}
	\frac{\rho_\phi}{\rho_\text{rad}} \sim \left( \frac{\alpha}{\lambda} \right)^2.
\end{align}
This indicates that 
the oscillation time scale of soft modes $(p < Q)$ always oscillates much slower than the typical time scale of interaction of $\chi$ 
with the thermal plasma, $Q \sim \lambda^2 \tilde \phi \lesssim (\lambda \alpha)^{1/2} T \ll \Gamma_{\text{th},\chi/\psi}$;
and that the turbulent evolution toward the UV regime is much slower than the typical interaction time scale of particles in thermal bath,
$t^{-1}_\text{quant} \sim ( \lambda^5 \alpha)^{1/2} T \lll \Gamma_{\text{th},\chi/\psi}$.
Correspondingly, the $\phi$ condensation dissipates much faster than the turbulent time scale driven by the four point interaction
\begin{align}
	\Gamma_\phi t_\text{quant} \gtrsim \lambda^4 T t_\text{quant} \gg 1.
\end{align}
Importantly, this implies that
before the quartic interaction completes the energy cascades toward the UV regime,
at least the interaction with the background thermal plasma dominates the UV cascade.
In addition, it is also possible that the quadratic term dominates the effective potential before the completion of the UV cascade.
Note that in the case of $\rho_\phi/ \rho_\text{rad} \lesssim (\alpha/\lambda)^2$ the above conditions are satisfied much easier.

Therefore, the $\phi$ particles produced by the four-point interaction is at most accumulated in the infrared regime
$p \ll \Gamma_{\text{th},\chi/\psi}$ 
and their evolution toward the UV regime is dominated by the interaction with the thermal plasma.
The first result implies that the $\chi$ particles see the soft $\phi$ as a slowly oscillating almost homogeneous background,
and hence the dissipation rate of $\phi$ with $p \ll \Gamma_{\text{th},\chi/\psi}$ can be approximated with that given in Sec.~\ref{sec:DTH}.
In the following section, we will see explicitly such produced $\phi$ particles obey the same equation with
that of condensation.

\subsection{Equation for $\phi$ particles}
\label{sec:ds}

In this section we derive an equation of motion of $\phi$ particles who are accumulated at low momenta. 
We assume the separation of time scales between thermal bath ($\Gamma_{\text{th},\chi}$) and $\phi$
particles: $\Gamma_{\text{th},\chi/\psi}\gg \Omega_{\phi,{\bf k}}$ where
$\Omega_{\bullet,{\bf k}} \equiv \sqrt{m_{\bullet,\text{eff}}^2 + {\bf k}^2}$ with $\bullet = \phi, \chi$.
We focus on the situation where $\chi$ particle is well thermalized.
In such a situation, 
one can use the quasi-particle ansatz for 2 point Green function of $\phi$. 
We assume the following quasiparticle form for the propagators:
 \beq
 G^{\chi}_{H}(t,t',{\bf k})&=&(1+2f_{B,{\bf k}}(t_c))\frac{\cos(\Omega_{\chi,{\bf k}}(t_c)t_{\Delta})}{\Omega_{\chi,{\bf k}}(t_c)}
 e^{\frac{-\Gamma_{\text{th},\chi}|t_{\Delta}|}{2}},\\
 iG^{\chi}_{J}(t,t',{\bf k})&=&\frac{\sin(\Omega_{\chi,{\bf k}}(t_c)t_{\Delta})}{\Omega_{\chi,{\bf k}}(t_c)}e^{\frac{-\Gamma_{\text{th},\chi}|t_{\Delta}|}{2}},\\
 G^{\phi}_{H}(t,t',{\bf k})&=&(1+2f_{\phi,{\bf k}}(t_c))
 \frac{\cos(\Omega_{\sigma,{\bf k}}t_{\Delta})}{\Omega_{\sigma,{\bf k}}}
,\\
 iG^{\phi}_{J}(t,t',{\bf k})&=&\frac{\sin(\Omega_{\phi,{\bf k}}t_{\Delta})}
 {\Omega_{\phi,{\bf k}}},
 \eeq
 where
 \beq
 t_c&\equiv&\frac{1}{2}(t+t'),\\
 t_{\Delta}&\equiv&t-t',
 \eeq
 and
 \beq
 G(x^0,y^0,{\bf k})=\int d^3({\bf x-y})e^{-i{\bf k}\cdot {\bf x-y}}G(x,y),
 \eeq
 and $f_{\rm B}$ is the Bose distribution function.
 The Kadanoff-Baym equations for $G^{\phi}_H$ is
 \footnote{If $\phi$ condensation
oscillates with thermal log potential, the number density of tachyonic mode of $\phi$ particles
may grow. However, the energy density of $\phi$ particles is at most that of condensation.
Therefore, we can neglect such effects.}
 \beq
 \label{a0}
 (\Box_x+m^2_{\phi}(x))G^{\phi}_H(x,y)=&i&\int_0^{x^0} d^4z \Pi_{J}^{\phi}(x,z)G^{\phi}_H(z,y)
 \nonumber \\
                                                                       &-&i\int_0^{y^0} d^4z \Pi_{H}^{\phi}(x,z)G^{\phi}_J(z,y).
 \eeq
 Thanks to the dumplings caused by $\Gamma_{\text{th},\chi}$, this non local equation can be regarded as
 local one in time scale of $\Omega_{\phi}$.
 Neglecting terms of $O({\Omega_\phi}/{\alpha T})$, one can get the following Boltzmann type 
 equations\footnote{To derive Eq.~(\ref{BLT}), we follow Ref.~\cite{Hohenegger:2008zk}.}
\begin{align}
\label{BLT}
	\frac{df_{\phi}(\Omega_{\bf k},t)}{dt}
	-\left[\frac{{\bf k}^2}{\Omega_{\bf k}}H-\frac{\dot T}{T}
	\frac{\frac{\partial m_{\phi}^2(T)}{\partial \ln T}}{2\Omega_{\bf k}}\right]
	\frac{\partial f_{\phi}(\Omega_{\bf k},t)}{\partial \Omega_{\bf k}}
	=-\Gamma_{\phi}^{\rm par}(\tilde{\phi}(t))\left[f_{\phi}(\Omega_{\bf k},t)-f_B({\bf k})\right],
\end{align}
where $\Gamma_{\phi}^{\rm par} (\tilde{\phi})$ is oscillation averaged dissipation rate defined as
\begin{align}
	\Gamma_{\phi}^{\rm par}(\tilde{\phi}(t)) \equiv \langle \Gamma_{\phi} [\phi(t)] \rangle.
\end{align}
One can see that the dissipation rate of particle $\Gamma_{\phi}^{\rm par}$ and that of condensation
(\ref{os_av_dec}) are the same order.
 %
 %
 %
 \section{Estimation of $k$}
 \label{app:ESTK}
 
 In this appendix, we give a formula for evaluation of $k$ which
is defined in (\ref{alwc}) with general set up.
We assume that 
$\bar{\phi}$, which is the amplitude of $\phi$, and
the total energy density other than $\phi$, which we denote by $\rho_{\rm oth}$, have
power low dependences on the scale factor as
\begin{align}
\bar{\phi}&\propto R^{-a},\\
\rho_{\rm oth}&\propto R^{-b}, 
\end{align}
with $R$ being the scale factor.\footnote{
	It should not be confused with the energy fraction of curvaton.
}
The power law index $a$ or $b$ may change at some epoch, depending on the form of the scalar potential
and the properties of the background.
The epoch of the transition of $a$ or $b$ can be written in the following form
\begin{align}
\rho_{\rm oth}^{\beta}{\bar{\phi}}^{\alpha} =M^{4\beta+\alpha},
\end{align}
with $\alpha$ and $\beta$ being some constants and $M$ being a mass scale, which is assumed to be a constant.
With these assumptions, we can estimate $k$.

To be more general, we assume that the power law index $a$ or $b$ change $N$ times
from the start of $\phi$ oscillation to its decay.
The condition of $n$-th transition $(1\leq n \leq N)$ can be written as 
\begin{align}
\label{cnd:kr}
	\rho_{\rm oth}^{\beta_n}{\bar{\phi}_n}^{\alpha_n}=M^{4\beta_n+\alpha_n}_n
\end{align}
We can define the time slicing, which satisfy the $n$-th transition condition and
we will call the surface $n$-th surface.
We set the condition of onset of the oscillation to be $\rho_{\rm oth}\bar{\phi}^{\alpha_0}=M_0^{4+\alpha_0}$
and call it $0$-th surface. 
We also take $N+1$-th surface at the time just before decay with the condition $\rho_{\rm oth}=M_{N+1}^4$.
The time evolution of $\bar{\phi}$ and $\rho_{\rm oth}$ from $n-1$ slicing to $n$ slicing
is assumed to be the following form
\begin{align}
	\bar{\phi}&\propto R^{-a_n},\\
\rho_{\rm oth}&\propto R^{-b_n}.
\end{align}
Now we have fixed all components needed to estimate $k$.
Suppose that at $n-1$-th surface $\bar{\phi}$ and $\rho_{\rm oth}$ are depend on $\phi_i$ as
\begin{align}
	\bar{\phi}_{n-1}&\propto \phi_i^{A_{n-1}},\\
\rho_{n-1,{\rm oth}}&\propto \phi_i^{B_{n-1}}.
\end{align}
Then, the subsequent evolution can be written as
\begin{align}
	\bar{\phi}&\propto \phi_i^{A_n}\left(\frac{R}{R_{n-1}}\right)^{-a_n},\\
\rho_{\rm oth}&\propto \phi_i^{B_n}\left(\frac{R}{R_{n-1}}\right)^{-b_n},
\end{align}
with $R_{n-1}$ being the scale factor at $n-1$-th slice.
Using the condition (\ref{cnd:kr}), one can obtain
\begin{align}
	\left(\frac{R_n}{R_{n-1}}\right)&\propto 
	\phi_i^{\frac{A_{n-1}\alpha_n+B_{n-1}\beta_n}{a_n\alpha_n+b_n\beta_n}}\equiv
	\phi_i^{C_n},\\
\bar{\phi}_{n}&\propto \phi_i^{A_{n-1}-a_nCn},\\
\rho_{n,{\rm oth}}&\propto \phi_i^{B_{n-1}-b_nCn}.
\end{align}
Thus, what we have to do is just to solve the following series
\begin{align}
	A_{n}&=A_{n-1}-a_nC_n,\\
	B_{n}&=B_{n-1}-b_nC_n,\\
	C_n&=\frac{A_{n-1}\alpha_n+B_{n-1}\beta_n}{a_n\alpha_n+b_n\beta_n},
\end{align}
with initial condition $A_0=1,B_0=-\alpha_0$.
Then, $k$ can be obtained as
\begin{align}
	k={k_aA_{N+1}+k_bB_{N+1}},
\end{align}
where we assume the form of energy density just before decay as
$\rho_\phi\propto \bar{\phi}^{k_a}\rho_{\rm oth}^{k_b}$. Note that
$3(1+w_\phi^{\text{(dec)}})=a_{N+1}k_a+b_{N+1}k_b$.
%
%
%
 %
 %
 %
 %


\end{document}